\documentclass[]{emulateapj}

\newcommand\coplus{CO$^+$~}

\newcommand\hcoplus{HCO$^+$~}

\newcommand\hocplus{HOC$^+$~}

\newcommand\hho{H$_2$O~}

\newcommand\hh{H$_2$~}

\newcommand{\tto}[2]{$#1 \times 10^{#2}$}

\shorttitle{Multidimensional chemical modelling, II. Irradiated outflow walls}
\shortauthors{S. Bruderer et al.}

\begin{document}

\title{Multidimensional chemical modelling of young stellar objects,\\ II. Irradiated outflow walls in a high-mass\\ star forming region}

\author{S. Bruderer\altaffilmark{1}}
\altaffiltext{1}{Institute of Astronomy, ETH Zurich, CH-8093 Zurich, Switzerland}
\email{simonbr@astro.phys.ethz.ch}

\author{A.O. Benz\altaffilmark{1}}

\author{S. D. Doty\altaffilmark{2}}
\altaffiltext{2}{Department of Physics and Astronomy, Denison University, Granville, OH  43023, USA}

\author{E. F. van Dishoeck\altaffilmark{3,4}} 
\altaffiltext{3}{Leiden Observatory, Leiden University, PO Box 9513, 2300 RA Leiden, The  Netherlands}
\altaffiltext{4}{Max-Planck Institut f\"ur Extraterrestrische Physik (MPE), Giessenbachstrasse 1, 85748 Garching, Germany}

\author{T. L. Bourke\altaffilmark{5}}
\altaffiltext{5}{Harvard-Smithsonian Center for Astrophysics, 60 Garden Street, Cambridge, MA 02138, USA}

\begin{abstract}
Observations of the high-mass star forming region AFGL 2591 reveal a large abundance of CO$^+$, a molecule known to be enhanced by far UV (FUV) and X-ray irradiation. In chemical models assuming a spherically symmetric envelope, the volume of gas irradiated by protostellar FUV radiation is very small due to the high extinction by dust. The abundance of \coplus is thus underpredicted by orders of magnitude. In a more realistic model, FUV photons can escape through an outflow region and irradiate gas at the border to the envelope. Thus, we introduce the first 2D axi-symmetric chemical model of the envelope of a high-mass star forming region to explain the \coplus observations as a prototypical FUV tracer. The model assumes an axi-symmetric power-law density structure with a cavity due to the outflow. The local FUV flux is calculated by a Monte Carlo radiative transfer code taking scattering on dust into account. A grid of precalculated chemical abundances, introduced in the first part of this series of papers, is used to quickly interpolate chemical abundances. This approach allows to calculate the temperature structure of the FUV heated outflow walls self-consistently with the chemistry.\\
Synthetic maps of the line flux are calculated using a raytracer code. Single-dish and interferometric observations are simulated and the model results are compared to published and new JCMT and SMA observations. The two-dimensional model of AFGL 2591 is able to reproduce the JCMT single-dish observations and also explains the non-detection by the SMA. We conclude that the observed \coplus line flux and its narrow width can be interpreted by emission from the warm and dense outflow walls irradiated by protostellar FUV radiation.
\end{abstract}


%
% Sec: Introduction
%
\section{Introduction} \label{sec:intro}

Forming high-mass O and B stars with a surface temperature larger than $\approx 10^4$ K emit most of their radiation in far UV (FUV) wavelengths. After a not yet well defined point of the evolution, also X-rays are emitted. When still deeply embedded in their natal cloud, most of this high-energy radiation is absorbed by the large column density of gas and dust towards the protostar. It is thus not available to direct observation but does influence the composition of the molecular envelope through heating and photoionization. An essential ingredient of star formation are bipolar outflows transporting the excess angular momentum outwards. When the fast, low-density gas of molecular outflows or jets expand into the surrouding molecular cloud, large cavities may result (e.g. \citealt{Preibisch03}). Along these outflow cavities, FUV radiation may escape and irradiate the high density material at the border of the cavity.\\

A surprisingly large amount of \coplus has been detected in the high-mass star forming region AFGL 2591 by \citet{Staeuber07}. Their radiative transfer calculations indicate fractional abundances of order $10^{-10}$, much higher than $10^{-15}-10^{-14}$ predicted by dark cloud models. \coplus has previously not been detected towards envelopes of young stellar objects (YSOs), but has been seen in photo dominated regions (PDRs, e.g. \citealt{Jansen95b} or \citealt{Fuente03}). Strong far UV (FUV) radiation in these regions heats the gas and drives a peculiar chemistry which enhances the abundance of CO$^+$. The detection of \coplus in envelopes of YSOs is thus strong evidence for the feedback of the protostar on the envelope by high-energy irradiation. In this work, we will study \coplus as a prototypical species enhanced by FUV irradiation in high-mass YSOs.\\

Chemical models solve for the evolution of the abundances of molecules and atoms. They simulate a network of species reacting with each other. The network consists of different types of chemical reactions. In dark clouds for example, all FUV radiation is shielded by a large column density of dust and the chemistry is dominated by cosmic ray ionization and reactions between neutral and ionized species (e.g. \citealt{Doty02}). In the innermost part of a YSO  envelope, X-rays may dominate the ionization (\citealt{Staeuber05}). They influence the chemistry mainly by secondary fast photoelectrons hitting molecules and atoms. X-ray induced chemistry is very similar to cosmic ray induced chemistry (\citealt{Bruderer09a}). In regions with a strong FUV irradiation, direct photoionization of species with ionization potential $< 13.6$~ eV drive the chemistry. The influence of FUV radiation on the envelope of YSOs has been studied by \citet{Staeuber04}.\\

So far, a spherically symmetric structure of the envelope has been assumed for chemical models of envelopes of high-mass star forming regions (e.g. \citealt{Viti99a}, \citealt{Doty02}, \citealt{Staeuber04, Staeuber05}). While the abundances derived using this spherical model agree for many species with observations of specific sources, they fail to explain the amount of CO$^+$. Due to the attenuation by dust, FUV radiation cannot escape the innermost few hundred AU and the volume of gas where \coplus may be formed is too small in spherical models to reproduce the observations. Another possible formation mechanism of \coplus is X-ray irradiation. Unlike FUV photons, X-rays are mainly attenuated by geometrical dilution ($\propto r^{-2}$) due to their smaller absorption cross-section proportional to $\lambda^3$. X-ray radiation can thus penetrate much deeper into the envelope on scales of a few thousand AU. Protostellar X-ray luminosities of up to $10^{32}$ erg s$^{-1}$ in the 1-100 keV band are observed (\citealt{Preibisch05a}). For this luminosity however, the X-ray flux is too low to enhance \coplus sufficiently to match with observations.\\

The interaction zone between outflow and infall has been proposed before as sites of anomalous chemistry and molecular excitation. \citet{Spaans95} have explained the strong CO(6-5) emission in narrow lines observed towards many low-mass YSOs by heating of the cavity wall through protostellar FUV photons. Compact emission of \hcoplus associated with the outflow wall has been detected by \citet{Hogerheijde98} in the Class 0 object L 1527. A possible explanation is mixing of partially ionized gas from the outflow with infalling material leading to a thin layer of dense and ionized gas, where species with a high dipole moment (e.g. \hcoplus or HCN) can be excitet efficiently by collisions with electrons (\citealt{Hogerheijde98b}). Detailed 3D radiative transfer studies by \citet{Rawlings04} on the other hand prefer a scenario with an enhanced abundance of \hcoplus due to shock liberation and photoprocessing of molecular material stored in ice mantles. Observational evidence for such an interaction layer in a high-mass star forming region are found by \citet{Codella06} by studying the line profiles of different molecular species. Recently, \cite{vanKempen09a} find evidence for UV photons escaping through the outflow cones and impacting the walls for the low-mass protostar HH 46.\\

In this work, we will introduce a chemical model that implements a non-spherical density structure taking an outflow cavity into account. A proper treatment of the FUV irradiated outflow walls requires two-dimensional chemical models with high spatial resolution. This is however computationally excessively  expensive, especially if the temperature structure is calculated self-consistently with the chemical abundances. In the first paper of this series (\citealt{Bruderer09a}), we have introduced a new method for fast chemical modelling. A precalculated grid of chemical abundances depending on different physical parameters like the density, temperature or FUV flux is calculated and the abundances can then be obtained by interpolation in this database. Using different benchmark tests, we found very good agreement between fully calculated chemical abundances and interpolated values, while the interpolation approach is more than 5 orders of magnitudes faster. This speed-up allows to quickly construct detailed chemical models with a complex geometry.\\

We construct a detailed two-dimensional chemical model of the high-mass star forming region AFGL 2591. The goal is to explain the high abundance of \coplus measured by the JCMT single-dish telescope (\citealt{Staeuber07}) as well as the non-detection by the Submillimeter Array (SMA) interferometer (\citealt{Benz07} and new data, reported in Sect. \ref{sec:compsma}). In this paper, we assume a fixed geometry and consider only CO$^+$. The next part of this series of papers is dedicated to a study of the influence of geometry for a larger sample of species.\\

The paper is organized as follows: In the first section of the paper, we briefly discuss the chemistry of CO$^+$. The next section introduces the two-dimensional model of AFGL 2591: We describe the modelling process, the radiative transfer of the FUV radiation and the calculation of the temperature structure. In section \ref{sec:results}, we present results of the chemical model. Modeled fluxes and synthetic maps are compared to JCMT and SMA observations.\\

%
% Subsec: Chemistry of CO+
%
\section{Chemistry of \coplus} \label{sec:chemcoplus}

\begin{figure*}[tbh]
\begin{center}
\includegraphics[width=0.85\textwidth]{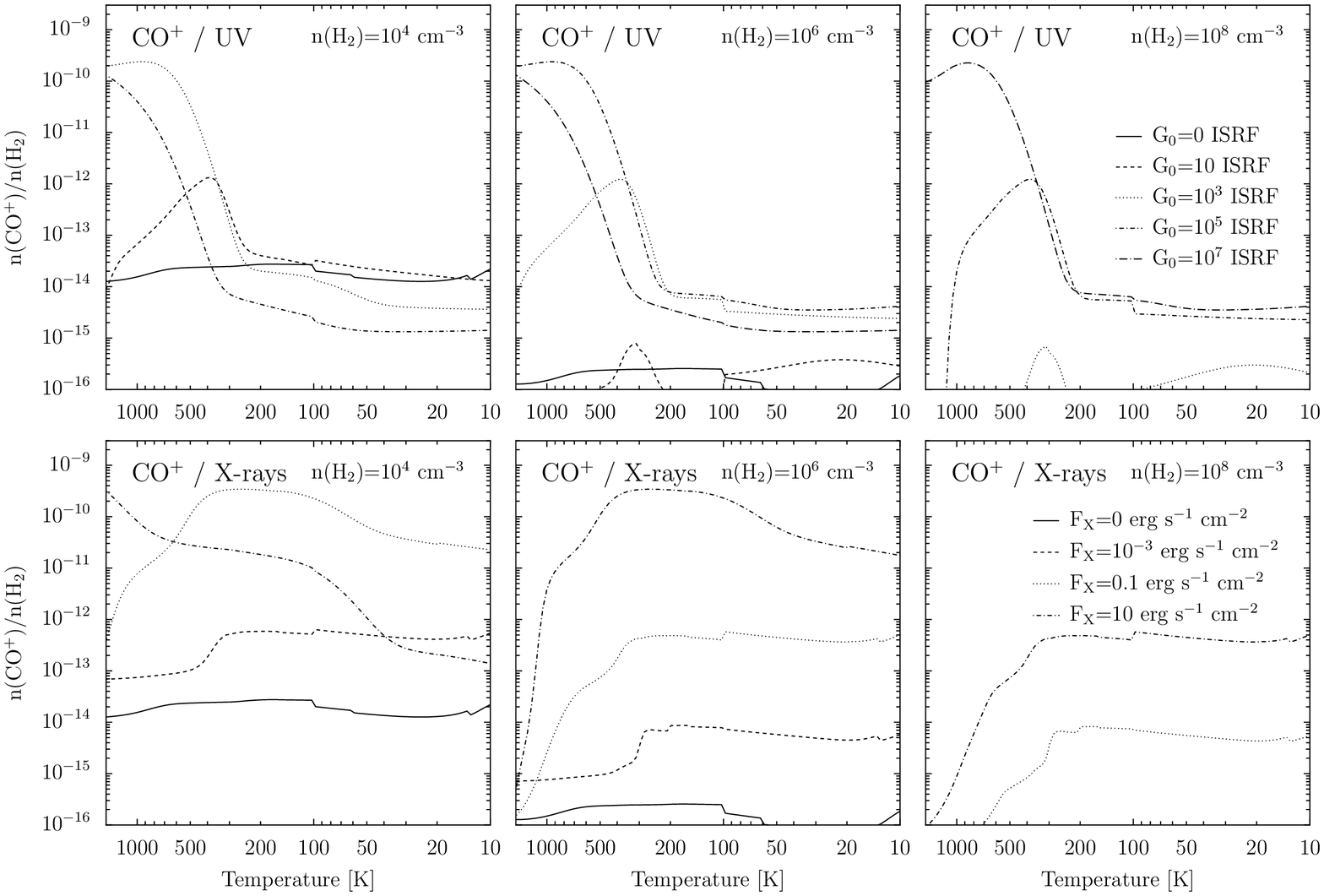}
\caption{Parameter study of \coplus: The fractional abundance is shown for a temperature range between 10 and 1500 K, different FUV fluxes between 0.1 and $10^7$ ISRF at $\tau=1$ in units of A$_V$ (top panels) and X-ray fluxes between $10^{-10}$ and 10 erg cm$^{-2}$ s$^{-1}$ for an attenuating hydrogen column density of $10^{20}$ cm$^{-2}$ (bottom panels). Plots for hydrogen densities of $10^4$, $10^6$ and $10^8$ cm$^{-3}$ are given. For comparision, observed abundances are of order $10^{-10}$.\label{fig:plot_param_co+}}
\end{center}
\end{figure*}

The fractional abundance of \coplus depending on the temperature is given in Fig. \ref{fig:plot_param_co+} for different gas densities, X-ray fluxes [erg s$^{-1}$ cm$^{-2}$] and FUV irradiation in units of the interstellar radiation field (ISRF). A parcel of gas has been modeled for this figure using the chemical model described in the first paper of this series (\citealt{Bruderer09a}). The chemical model solves for the evolution of the abundances of a network of chemical species starting from initial abundances. We assume a chemical age of \tto{5}{4} yr as suggested by \citet{Staeuber05}. Chemical time-scales are very short in regions with strong FUV or X-ray irradiation, where a significant amount of \coplus is produced. Equilibrium conditions are thus reached very quickly and the chemical age does not influence the resulting abundances. Photodissociation and ionization rates depend on the attenuating column density between the FUV/X-ray emission and the modeled parcel of gas. We assume an optical depth of $\tau=1$ (FUV) and a hydrogen column density of $10^{20}$ cm$^{-2}$ (X-rays) to derive the irradiating spectra.\\

In gas with strong FUV irradiation, \coplus is efficiently produced by the reaction C$^+$ + OH $\rightarrow$ \coplus + H (\citealt{Staeuber07}, \citealt{Fuente06}). Direct photoionization of carbon atoms accounts for C$^+$, while OH is produced by the photodissociation of water. At temperatures below 100 K, \hho is frozen-out and the amount of OH reduced. Between 100 K and about 250 K, the reaction O + \hh $\rightarrow$ OH + H does not proceed due to an activation barrier and OH is destroyed through photodissociation. The amount of OH increases with higher temperature and thus \coplus is produced. At very high FUV fields, \coplus is photodissociated and its abundance again decreases.\\

In X-ray irradiated gas, ionization of CO through secondary electrons accounts for the production of CO$^+$. The abundance of \coplus is thus constant with density and the fractional abundance $\sim$ inversely proportional to the total density. In low density gas with high X-ray irradiation, the production of \coplus through C$^+$ + OH can also be important as Figure \ref{fig:plot_param_co+}  shows at a density of 10$^4$~ cm$^{-3}$ and an X-ray flux of 10 erg cm$^{-2}$ s$^{-1}$. In this regime, the fractional abundance of \coplus increases with temperature.\\

\coplus is destroyed by \hh leading to \hocplus and \hcoplus with equal branching ratios. In regions with very high FUV irradiation, H$_2$ is photodissociated to atomic hydrogen and \coplus is destroyed by the reaction \coplus + H $\rightarrow$ CO + H$^+$.\\

%
% Figure plot_param_co+
%

To produce the observed fractional abundances of order $10^{-10}$ (\citealt{Staeuber07}) by X-rays, a flux higher than 10 erg cm$^{-2}$ s$^{-1}$ is needed for a density of 10$^6$ cm$^{-3}$. For an X-ray luminosity of 10$^{32}$ erg s$^{-1}$ as found by \citet{Staeuber05} for AFGL 2591 and in the extreme case of no attenuation, this flux can only be achieved in the innermost 60 AU. In a spherical model of AFGL 2591, X-rays may enhance the abundance of \coplus up to a few times $10^{-13}$, and \citet{Staeuber05} have thus suggested the molecule as a tracer for X-rays. X-rays are however not able to enhance the abundance to $10^{-10}$ on larger distances to the source. The FUV flux of a young B star on the other hand is still sufficient at a distance of almost 10\,000 AU to produce a fractional abundance exceeding 10$^{-10}$ if no dust attenuates the FUV field. We thus consider \coplus to be a tracer for FUV irradiation rather than for X-rays.\\

A short note concerning the combined influence of X-rays and FUV radiation: in the vicinity of a strong FUV source, CO is dissociated and X-rays cannot produce \coplus efficiently. For example, the calculation of the parameter study including X-rays at a density of 10$^4$~ cm$^{-3}$ has been repeated in the presence of an FUV field of 10$^4$ ISRF. For this condition, the fractional abundance does not exceed 10$^{-14}$ and we thus expect the fractional abundance in a low density and X-ray/FUV irradiated outflow region to be very low.

\section{An axi-symmetric model of AFGL 2591} \label{sec:afgl2dmod}

\begin{figure}[ht]
\plotone{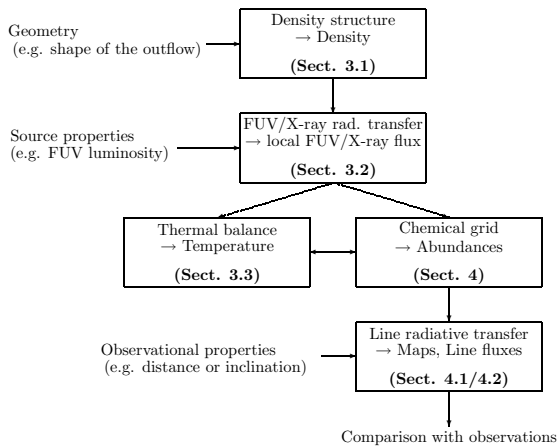}
\caption{Flowchart of the modelling process for the axi-symmetric model of AFGL 2591.\label{fig:flowchart}}
\end{figure}

Figure \ref{fig:flowchart} summarizes the process for an axi-symmetric model of AFGL 2591. We first assume a density structure in Sect. \ref{sec:dens2dafgl}. This density distribution is used in the following section to calculate the local FUV and X-ray field at every position of the model. The high-energy flux together with the density structure then enters the thermal balance calculation, which solves for the temperature structure. Since cooling and heating depend upon the composition of the gas, the temperature must thus be calculated self-consistently with the chemical abundances which are read out of the grid of chemical models presented by \citet{Bruderer09a}. The abundance of \coplus is then obtained using the same approach. Synthetic maps are calculated using a raytracer and convolved to the appropriate telescope beam for comparison with single-dish observations. To compare model results to interferometric observations, the synthetic maps are converted into visibility amplitudes and reduced in the same way as interferometric observations.

%
% Figure fluss.eps
%

%
% Subsec: Physical structure
%
\subsection{Density structure} \label{sec:dens2dafgl}

\begin{figure*}[tbh]
\includegraphics[width=\textwidth]{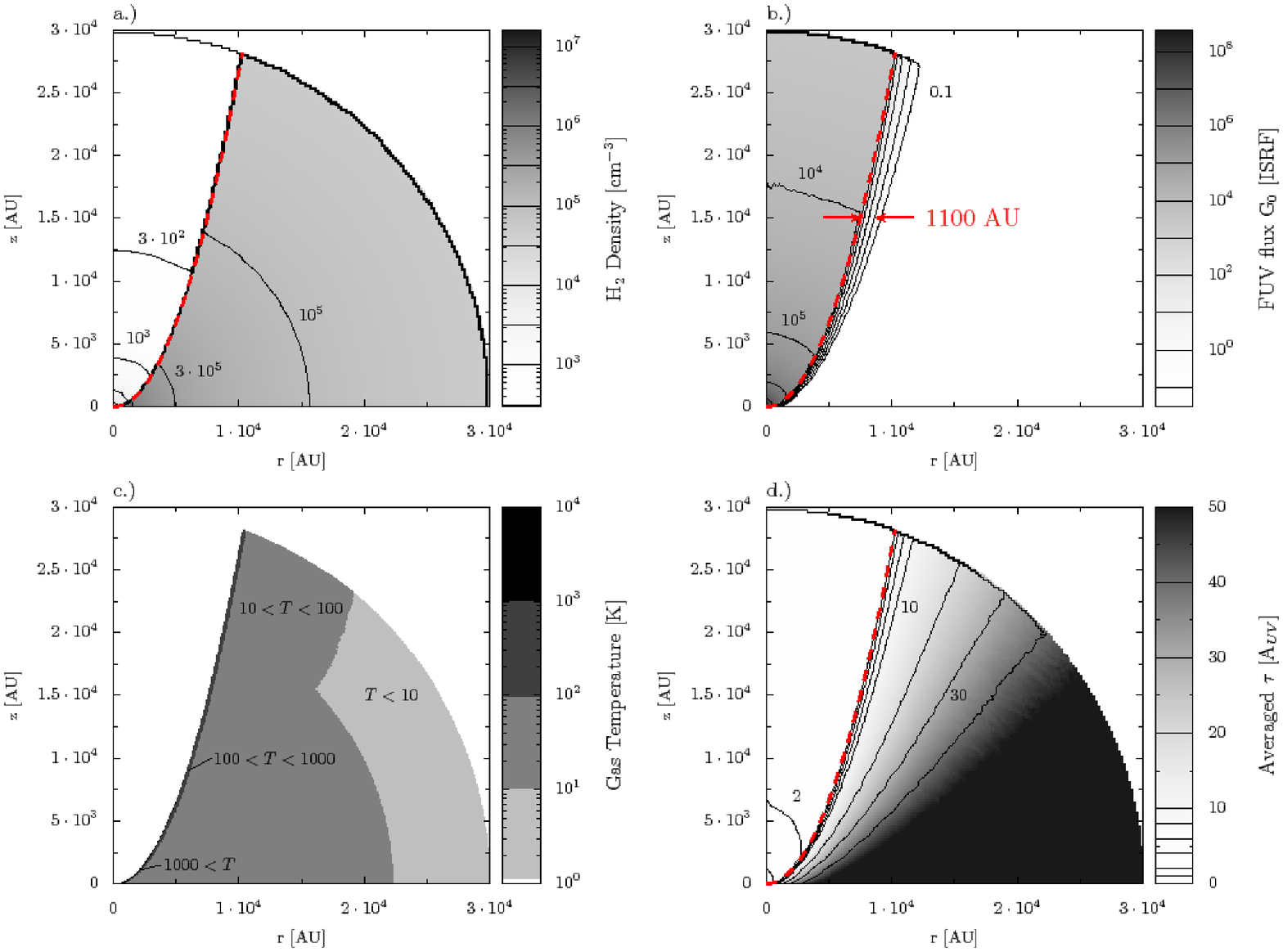}
\caption{Physical structure of the (axi-symmetric) 2D model of AFGL 2591: \textbf{a.)} density structure (Sect. \ref{sec:dens2dafgl}). The border between the outflow and infall-region is indicated by the dotted line. \textbf{b.)} Attenuated FUV flux calculated by the FUV radiative transfer code (Sect. \ref{sec:rad2dafgl}). The gray/red arrow indicates the width of the outflow wall until $G_0$ is attenuated to 1 ISRF. \textbf{c.)} Temperature structure (Sect. \ref{sec:temp2dafgl}) \textbf{d.)} Averaged value of $\langle \tau \rangle$ in units of A$_{\rm UV}$ (Sect. \ref{sec:rad2dafgl}).\label{fig:out_2d_flared}}
\end{figure*}

One goal of this work is to study the difference in the modeled \coplus line strength using a spherically symmetric 1D and an axi-symmetric 2D model. For the 2D model we thus assume a density distribution following the power-law of the spherically symmetric 1D model, except for an outflow-region (Fig. \ref{fig:out_2d_flared}a). We implement $n({\rm H}_2) = n_0 (r_0/r)$ with $n_0=5.8 \times 10^4$ cm$^{-3}$ and $r_0=2.7 \times 10^4$ AU, as proposed by \citet{vdTak99}.\\

For the separation between the outflow region and the envelope, we use the function
\begin{equation}
z = \left( \frac{1}{10\,000 \tan^2\left( \alpha / 2 \right)}  \right)  \times r^2 \ \ ,
\end{equation}
where $z$ and $r$ denote coordinates along the outflow axis and perpendicular to the outflow (in units of AU). The full opening angle $\alpha$ at $z=10\,000$~ AU is assumed to be $\approx 62^\circ$, approximately in agreement with the high resolution mid-IR observations by \citet{Preibisch03}. The full opening angle of the outflow cone at $z=30\,000$ AU is about 40 degrees. The choice of a parabolic outflow cavity is supported by theoretical predictions (\citealt{Canto08}) and observations (\citealt{Velusamy98}). It is also compatible with the mid-IR observations of AFGL 2591.\\

Vibrationally excited CO at a velocity of $-200$ km s$^{-1}$, emitted in outflow gas has been observed by \citet{vdTak99}. Assuming an infall velocity in the envelope of order 10 km s$^{-1}$, pressure equilibrium ($\rho_{\rm in} v_{\rm in}^2 = \rho_{\rm out} v_{\rm out}^2$) requires a density ratio of \tto{2.5}{-3} between the two regions. In the outflow region, the density of the power-law distribution is thus assumed to be reduced by this ratio. As in the spherical models of Doty et al. and St\"auber et al., we do not model the innermost $\approx$250 AU. We make use of the symmetry and only model positive values of $r$ and $z$. In this quadrant, the model consists of 300 cells along the outflow ($z$-axis) and 450 cells parallel to the midplane ($r$-axis) thus a total of $135\,000$ cells. The resolution is thus about 100 AU $\times$ 67 AU.\\

%
% Figure out_2d_flared
%

%
% Subsec: High energy radiation
%
\subsection{FUV and X-ray radiative transfer} \label{sec:rad2dafgl}

High-energy radiation with photon energy in the range of 6 - 13.6 eV (FUV radiation) and 0.1 - 100 keV (X-rays) penetrates molecular gas, ionizes molecules and heats the gas. The local FUV and X-ray field at every position are thus required to calculate the temperature structure and the chemical abundances.\\

For the local X-ray intensity, we take into account the distance $r$ and the total hydrogen column density $N({\rm H}_{\rm tot})$ [cm$^{-2}$] between the X-ray source and the local position. For an X-ray luminosity $L_{\rm X}$ [erg s$^{-1}$], the local intensity is thus proportional to $L_{\rm X}/4\pi r^2 \times \exp\left(-\sigma_{\rm photo}(E) \cdot N({\rm H}_{\rm tot}) \right)$, where $\sigma_{\rm photo}(E)$ [cm$^{2}$] is the photoionization cross-section at a photon energy $E$ [eV]. We follow \citet{Staeuber05} and assume an X-ray emission by a thermal plasma at a temperature of \tto{7}{7} K and a protostellar X-ray luminosity of 10$^{32}$ erg s$^{-1}$. In their spherically symmetric models, the attenuating column density in the innermost 250 AU is a free parameter found to be about \tto{3}{22} cm$^{-2}$. For the 2D model, the same value is adopted in the midplane ($z=0$). Since density in the outflow is a factor of 400 lower, the column density between source and outflow walls can be of order 10$^{20}$ cm$^{-2}$, and photons with energies between 0.1 and 1 keV are not absorbed. Unlike \citet{Staeuber05}, we thus take into account these photons in the 2D model.\\

The central FUV source is assumed to emit a blackbody spectrum with a temperature of \tto{3}{4} K and a luminosity of \tto{2}{4} L$_\odot$ (\citealt{Staeuber04}). At a distance of 200 AU, the FUV flux in absence of any attenuation corresponds to \tto{2.3}{8} times the ISRF. Unlike elastic compton scattering for X-rays (\citealt{Staeuber05}), scattering of FUV photons on dust grains can be important due to the similar wavelength of FUV radiation and the dust size. A Monte Carlo simulation in the vein of \citet{vanZadelhoff03} calculates the local FUV strength (Appendix \ref{sec:mc_uvimplement}). Photon packages starting at the central source are traced through the density model, while scattering and attenuation by dust is taken into account. For simplicity, we do not calculate the FUV spectra explicitly, but deal with the scaling factor $G_0$ in units of the ISRF. To read out chemical abundances from the grid of chemical models, the visual extinction $\tau$ [A$_V$] is needed. We obtain the local extinction from 
\begin{equation} \label{eq:tauavg}
\langle \tau \rangle = - \log\left(\frac{\langle I^i_{\rm att} \rangle_i}{\langle I^i_{\rm unatt} \rangle_i}\right) \ \ ,
\end{equation}
where $I^i_{\rm att}$ is the local flux for a photon package $i$. The unattenuated flux $I^i_{\rm unatt}$ is only geometrically diluted. Brackets denote averaging over all photon packages passing through a cell. We implement the cross-section on dust given in \citet{Draine03a} for average Milky Way dust with $R_V$=3.1 and C/H=56 ppm in PAHs. This choice however has a minor effect on the model results, as photionization no longer dominates in regions, where a higher value of $R_V \approx 5.5$ is expected. Since the adopted approach corresponds to the calculation at one FUV wavelength, we use the dust properties at a photon energy of 9.8 eV, in the middle of the 6 - 13.6 eV FUV band. For the conversion $N({\rm H}_{\rm tot}) / A_V = 1.87 \times 10^{21}$~ cm$^{-2}$ (\citealt{Bohlin78}) and an extinction cross-section of \tto{1.29}{-21} cm$^{2}$ per H-atom at 9.8 eV, we obtain a conversion factor between the FUV and visual extinction of $A_{\rm UV} / A_V = 2.4$.\\ 

For the density profile of Sect. \ref{sec:dens2dafgl}, the calculated FUV flux $\langle I_{\rm att} \rangle$ and extinction $\langle \tau_{\rm UV} \rangle$ are given in Fig. \ref{fig:out_2d_flared}b and Fig. \ref{fig:out_2d_flared}d. The attenuated flux shows the region where FUV radiation influences the chemistry, as this value enters the heating rates and the photodissociation rates of the form $k=G_0 \cdot C \cdot  \exp\left(-\gamma \cdot \tau_{\rm v}\right)$, with $C$ and $\gamma$ fitting constants (e.g. \citealt{vanDishoeck88}) and $G_0$ the FUV flux in units of the ISRF. Note that the unattenuated flux $\langle I_{\rm unatt} \rangle$ has to be used together with the attenuation $\tau_{\rm v}$ to read out the abundances from the chemical grid.\\

\subsection{Temperature calculation} \label{sec:temp2dafgl}

Traditional PDR models (e.g. \citealt{Tielens85}) find the photoelectric effect of FUV photons on small dust grains and PAHs to be the dominant heating process in regions with strong FUV irradiation. In the dense  outflow walls of the 2D model, this process can easily heat the gas to temperatures above 250 K. For temperatures between 250 K and several hundred K, the fractional abundance of \coplus is a strong function of temperature (Sect. \ref{sec:chemcoplus}). It is thus important to carry out a detailed calculation of the temperature profile in the outflow wall. Input parameters are the gas density $n$ [cm$^{-3}$], the FUV field given by $G_0$ [ISRF] and $\tau$ [A$_V$] and the energy deposition per density by X-ray photons H$_{\rm X}$ [erg s$^{-1}$].\\

To obtain the gas temperature $T_{\rm gas}$, the thermal balance has to be solved. We assume steady state conditions. Thus the equilibrium between the total heating ($\Gamma_{\rm tot}$) and cooling ($\Lambda_{\rm tot}$) rate yields the temperature. Different physical processes contribute to the individual rates $\Gamma_i$ and $\Lambda_i$, where $i$ denotes processes. The inner energy per parcel of gas $\epsilon$ [erg cm$^{-3}$] is given by 
\begin{eqnarray*}
&{}&\frac{\delta \epsilon}{\delta t} \equiv 0 = \Gamma_{\rm tot} - \Lambda_{\rm tot} \\
&{}&=\sum_i \Gamma_i(n,G_0,\tau,H_{\rm X}, x_{{\rm H}_2},\ldots,N({\rm CO}),\ldots,T_{\rm gas},T_{\rm dust})\\
&{}&-\sum_i \Lambda_i(n,G_0,\tau,H_{\rm X}, x_{{\rm H}_2},\ldots,N({\rm CO}),\ldots,T_{\rm gas},T_{\rm dust}) \ \ .
\end{eqnarray*}
Important examples for heating processes are H$_2$ collisional de-excitation following pumping through FUV photons and the photoelectric effect. The cooling rate consists of gas-grain collisional cooling and atomic/molecular line emission (e.g. the [CII] line at 158 $\mu$m). An overview of the implemented cooling and heating processes is given in Appendix \ref{sec:app_thermal}.\\

Heating and cooling rates depend on the composition of the gas. For example the C$^+$ atomic line cooling depends on the fractional abundance $x_{{\rm C}^+}$ of ionized carbon. To obtain a self-consistent solution, the composition of the gas is read out from the grid of chemical models. Because the temperature also enters as a parameter for the interpolation of the fractional abundances, the equation for the thermal balance has to be solved iteratively.\\

\begin{figure}[ht]
\plotone{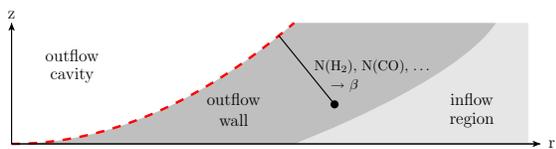}
\caption{Schematic overview on the 2D model. In the outflow wall, the simplified radiative transfer for heating and cooling is calculated between between the local position and the closest point of the outflow region. The column density (e.g. $N({\rm CO})$ or $N({\rm H}_2)$) along this line is used to calculate the escape probability $\beta$ of a photon.\label{fig:plot_1d1d}}
\end{figure}

Molecular or atomic line emission contributes to the cooling only if the photons can escape. A radiative transfer calculation is thus required to obtain proper cooling rates. For the atomic fine structure lines, we implement an escape probability code. As a simplification, the radiative transfer is not carried out in full 2D, but from the modeled point to the closest point of the outflow wall (Fig. \ref{fig:plot_1d1d}). The probability $\beta$ for a photon to escape from the outflow wall to the cavity is thus obtained from the molecular/atomic column density (e.g. $N({\rm CO})$ for CO) along this line.\\

%
% Figure plot_inclination
%

A fully self-consistent temperature model requires a two-dimensional radiative transfer calculation of the dust continuum radiation to obtain the dust temperature $T_{\rm dust}$. As a simplification, we use the dust temperature of the spherically symmetric model of \citet{Doty02} for the region where the FUV radiative transfer calculation predicts an optical depth larger than 10. In regions with $\tau_V < 10$, the dust temperature is obtained from the analytical expression given by \citet{Hollenbach91}. Using this approximation, the temperature structure remains consistent with the spherically symmetric model for temperatures below 100 K. In the particular case of CO$^+$, the relevant temperature range is $T > 250$~ K, and the interpolation to the spherically symmetric temperature profile is thus unimportant.\\

%
% Subsec: Results
%
\section{Results} \label{sec:results}

\subsection{Directly irradiated outflow walls} \label{sec:outflowwalls}

\begin{figure}[ht]
\plotone{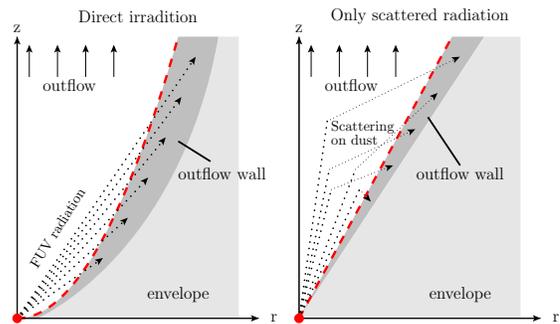}
\caption{Schematic view of the axi-symmetric model with conical and concave outflow cavity.\label{fig:fuvcomic}}
\end{figure}

The adopted separation between envelope and outflow ($z \propto r^2$) yields a ``flared'' geometry that allows direct irradiation of the dense outflow walls by the central FUV source. This results in a much larger FUV irradiated volume compared to a ``linear'' separation ($z \propto r$) where FUV photons can only penetrate the outflow walls if they are scattered on dust in the outflow (Fig. \ref{fig:fuvcomic}). Scattering is however relatively inefficient.\\

%
% Figure plot_co+_cut_1e32
%

What is the size scale of the outflow walls? We define the width of the outflow wall by the region with $G_0 > 1$ along a cut of constant $z$. For example at $z=15\,000$ AU, this corresponds to a thickness of about 1\,130  AU. The outflow cavity thus allows long-range streaming of FUV radiation to large distances from the central source. Photon packages do not necessarily follow cuts of constant $z$. The straight ray path from the FUV source to the far end of the outflow wall at $z=15\,000$ AU propagates for $9\,000$ AU in the outflow wall. Scattering thus extends the outflow wall. For comparison, the Monte Carlo code was re-run with scattering switched off. This run yielded a width of the outflow wall of only about 470 AU at $z=15\,000$, about 2.5 times smaller than the width calculated including scattering effects. This demonstrates the importance of scattering for the strengths of the local FUV field.\\

\begin{figure}[ht]
\plotone{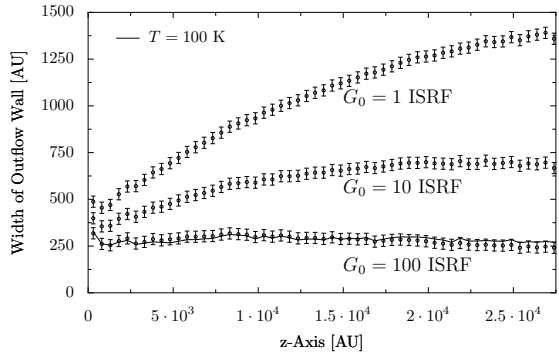}
\caption{Width of the outflow wall depending on the distance along the outflow axis. The figure shows the width defined by the region with temperature $>100 $ K (line) and with FUV flux larger than 1, 10 or 100 times the ISRF (error bars).\label{fig:width_outflow}}
\end{figure}

The width of the outflow wall with $G_0 > 1$, 10 or 100 ISRF depending on $z$ is given in Fig. \ref{fig:width_outflow} along with the error due to the finite resolution of the model. An FUV-enhanced layer along the outflow with a width of a few hundered AU is thus produced by the protostellar FUV radiation. While the width defined by $G_0=1$ increases with $z$, the layer with $G_0 > 100$ has an approximately constant width of $280 \pm 30$ AU. This is a consequence of the adopted geometry: For the $G_0 > 100$ layer, scattering is less important than for the $G_0 = 1$ region. Unscattered radiation at the surface layer of the outflow wall travels a longer way in the high density region for higher values of $z$ due to the larger impact angle. This effect compensates with the flux at the surface of the outflow wall $\propto 1/d^2$, where $d$ is the distance from the FUV source to the surface of the outflow wall, and the attenuation decreasing with $1/d$ for the implemented density profile.

%
% Figure width_outflow
%

Only a small fraction of the envelope in a thin layer on the outflow wall is heated by FUV photons to temperatures above 100 K. The temperature along cuts of fixed values of $z$ is shown in Fig. \ref{fig:plot_co+_cut_1e32}. For $z < 5000$ AU, the temperature at the surface exceeds 1000 K, consistent with the surface temperatures obtained from PDR models for similar conditions (e.g. \citealt{Kaufman99}).\\

The width of the region with $T> 100$ K in the outflow wall is approximately constant with $z$ (Fig. \ref{fig:width_outflow}) as the $G_0 = 100$ layer. In the region with $T \approx 100$ K, the heating is dominated by the photoelectric effect $\Gamma_{PE} \propto n \cdot G_0$, with $n$ being the density. The main coolant is atomic oxygen which is approximately thermalized and the cooling rate thus $\Lambda_{O} \propto n$. As found in the calculation, the gas temperature is thus approximately independent of density and thus follows $G_0$.\\

An overview of the mass and volume of different regions of the axi-symmetric 2D model compared to the spherical 1D model is given in Table \ref{tab:vol_afgl}. Indeed, the volume heated to $T > 100$~ K is more than two orders of magnitude larger in the 2D model, while the mass of the same region is only $\approx$ 8 times larger due to the adopted density gradient. Similarly, the volume irradiated by $G_0 > 1$ ISRF is almost 4 orders of magnitude larger in the 2D model, while the mass of the gas is about 2 orders of magnitude larger. We note that size of the region with $G_0 > 1$ ISRF in the spherical 1D model heavily depends on FUV attenuating column density in the innermost 250 AU. For Table \ref{tab:vol_afgl} the extreme case of no attenuation between the protostar and a radius of 250 AU was assumed.\\

Physical parameters like the cavity shape or the density ratio between infalling envelope and outflowing gas enter the modelling. A complete study is beyond the scope of this study and will be presented in the third paper of this series (Bruderer et al., in prep.). Here we discuss some general trends. A large surface directly irradiated by protostellar FUV radiation is required for a significant volume and mass of the dense and FUV heated material in the outflow wall. By changing the outflow angle $\alpha$ by $\pm 20$ degrees or the separation between outflow region and envelope to $z \propto r^{\alpha}$ with $\alpha=1.5 - 2.5$, the volume and mass with $T > 100$~ K vary only by a factor of about two. On the other hand, a higher density in the outflow region affects the model much more due to the higher extinction of FUV radiation. Increasing the density ratio $n_{\rm out} / n_{\rm in}$ from \tto{2.5}{-3} to $10^{-2}$ reduces the mass of the $T > 100$~ K region by a factor of 4, while the volume decreases by about an order of magnitude. These results for the relatively insensitivity with respect to angle and shape, and relative sensitivity to density, underline one of the key results of this paper -- namely the importance of direct irradiation on the chemical composition of outflow cavity walls.\\

\begin{table*}
\begin{center}
\caption{Volume and mass of different regions of the axi-symmetric model compared to the spherically symmetric model.\label{tab:vol_afgl}}
\begin{tabular}{l c c}
\tableline\tableline
Region                           & Mass       & Volume \\
                                 & [M$_\odot$] & [AU$^3$] \\
\tableline
\multicolumn{3}{l}{Axi-symmetric 2D model:} \\
Outflow                               & 0.014      & $1.1 \cdot 10^{13}$ \\
Envelope                              & 44.7       & $1.0 \cdot 10^{14}$ \\
Envelope ($T>100$ K)                  & 0.53       & $8.2 \cdot 10^{11}$ \\
Envelope ($G_0>1$)                    & 1.65       & $3.0 \cdot 10^{12}$ \\
Envelope ($G_0>1$ and $T<100$ K)      & 1.1        & $2.2 \cdot 10^{12}$ \\
\tableline
\multicolumn{3}{l}{Spherical 1D model:} \\
Envelope                              & 50.3       & $1.1 \cdot 10^{14}$ \\
Envelope ($T>100$ K)                  & 0.07       & $6.5 \cdot 10^{9}$ \\
Envelope ($G_0>1$)                    & 0.01       & $4.8 \cdot 10^{8}$ \\
\tableline
\end{tabular}
\end{center}
\end{table*}

The calculation of the FUV flux and gas temperature in the axi-symmetric 2D model suggests a region with temperature $T < 100$ K but FUV irradiation $G_0 > 1$ of potentially FUV irradiated ices. While a detailed discussion of these ices is beyond the scope of this paper, it is interesting to note two things. First, such a region exists due to the direct irradiation allowed by the concave outflow walls. Second, the potential mass in such a region is 1.1 M$_\odot$. Together, these facts suggest that further exploration of irradiated ices is warranted.\\

\subsection{\coplus abundance} \label{sec:coplusabundance}

Given the physical parameters for an axi-symmetric 2D model of AFGL 2591 defined previously, the grid of chemical models (\citealt{Bruderer09a}) can now be used to quickly read out the abundances of CO$^+$. Since no outflow features have been observed in lines of \coplus by \citet{Staeuber07}, we assume there is no \coplus in the outflow cavity. Also, the discussion in Sect. \ref{sec:chemcoplus} indicates a very low fractional abundance of \coplus in the outflow.\\

\begin{figure*}[tbh]
\includegraphics[width=\textwidth]{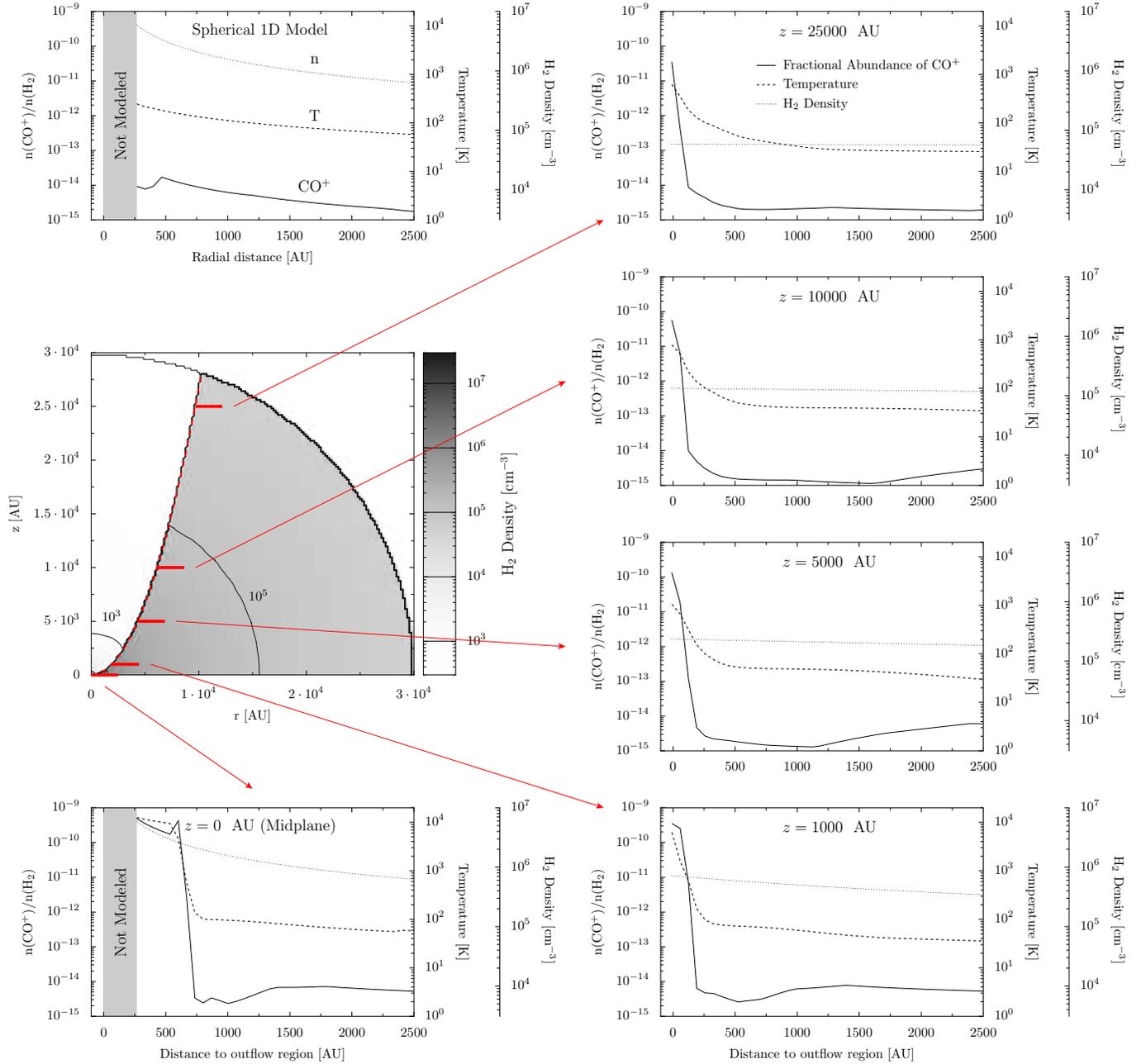}
\caption{\coplus in the two-dimensional axi-symmetric model of AFGL 2591. Slices at different values of $z$ are shown. The x-axis corresponds to the distance along the cut to the outflow region. The plots give the fractional abundance $x({\rm CO}^+)=n({\rm CO}^+)/n({\rm H}_2)$, the temperature and the density in solid, dashed and dotted lines, respectively. The figure in the top-left shows a spherical model of the same source. The gray rectangle indicates the innermost 250 AU where no modelling has been performed.\label{fig:plot_co+_cut_1e32}}
\end{figure*}

The fractional abundance of \coplus is given in Fig. \ref{fig:plot_co+_cut_1e32} along with the temperature and the density for cuts parallel to the midplane of the model. In a thin layer at the outflow wall, \coplus is enhanced. This layer matches with the FUV heated gas at temperatures above 250 K. The \coplus layer in the midplane ($z=0$) is considerably larger due to the scattering of FUV radiation from both outflows along the positive and negative z-axis. At the surface of the outflow wall, the fractional abundance of \coplus exceeds $10^{-10}$ for values of $z$ up to a few thousand AU. At $z=25\,000$, the fractional abundance of \coplus is still a few times $10^{-11}$.\\

Since the chemical time-scales of \coplus in regions with abundance larger than a few times $10^{-11}$ are shorter than 100 yrs, the evolutionary age of the source does only affect the modelling results if the FUV production of the protostar has not set in yet or the outflow cavity is not large enough yet. While direct evidence for protostellar FUV radiation of the source cannot be given, the mid-IR observations of the source by \citet{Preibisch03} suggest an outflow cavity with size of at least $10\,000$ AU along the outflow axis. For the measured flow velocity of a few hundered km s$^{-1}$ (\citealt{vdTak99}), the expansion to $10\,000$ AU would require $<500$ yrs much less than the infered age since YSO formation estimated to be a few times $10^{5}$ yrs (\citealt{Doty02,Doty06}).\\

For comparison, results of the spherical 1D model are given in the top left of Fig. \ref{fig:plot_co+_cut_1e32}. In this model, FUV radiation of the central source cannot escape the innermost part and the abundance of \coplus is thus very small at fractional abundances below $10^{-13}$. Outflow walls in the 2D model increase the surface irradiated by FUV and thus enhance the total amount of CO$^+$.\\

%
% Figure plot_co+_cut_1e32
%

Mixing between warm and ionized outflow material with the envelope can also induce a particular chemistry in a thin layer along the outflow. Charge-exchange reactions in this mixing layer enhance the abundance of CO$^+$. The width of the mixing layer is however small due to the high electron abundance leading to short recombination time-scales. While a detailed study (\citealt{Taylor95}, \citealt{Lim99} or \citealt{Nguyen00}) is beyond the scope of this work, a simple model is given in Appendix \ref{sec:coplusmix}. It predicts the amount of \coplus produced in this mixing layer to be enhanced, however insufficiently to explain observations.

%
% Subsec: Results
%
\subsection{Comparison with JCMT observations} \label{sec:compobs}

Two rotational lines of \coplus with $\Delta {\rm v} \approx 4$ km s$^{-1}$ have been detected by \citet{Staeuber07} at the center position of AFGL 2591 using the JCMT. We follow their arguments and assume the lines to be detected. The observed line fluxes are given in Table \ref{tab:coplus_molec} along with molecular data of \coplus taken from the JPL database (\citealt{Pickett98}). The Einstein-$A$ coefficients are scaled to a dipole moment of 2.77 D (\citealt{Rosmus82}).\\

\begin{table*}
\begin{center}
\caption{Molecular data of \coplus and integrated line flux at the center position of AFGL 2591 observed by the JCMT (\citealt{Staeuber07}).\label{tab:coplus_molec}}
\centering
\begin{tabular}{c c c c cc}
\tableline\tableline
Transition                        & Frequency   & $A_{\rm ul}$ &  $E_{\rm up}$ & $g_{\rm u}$ &  Line Flux  \\
                                  & [GHz]       & [10$^{-3}$ s$^{-1}$] & [K] &         & [K km s$^{-1}$]  \\
\tableline
$3\frac{5}{2}$ - $2\frac{3}{2}$   & 353.741     & 1.58            & 33.94   & 6    & 0.5      \\
$3\frac{7}{2}$ - $2\frac{5}{2}$   & 354.014     &  1.7           & 33.99    &  8  & 0.27     \\
\tableline
\end{tabular}
\end{center}
\end{table*}

Synthetic spectra are calculated to compare the abundance modelling results to these observations. The excitation of \coplus is needed to model the line flux. However, the mechanism is unclear since \coplus is more more likely to be destroyed than excited in collision with H$_2$. However, electrons and atomic hydrogen are abundant at the surface of PDRs and can excite \coplus as discussed in \citet{Andersson08}. Possibly, \coplus is formed in an excited state. In this work, we follow \citet{Staeuber07} and assume a fixed excitation temperature $T_{\rm ex}$. The effects of collisional excitation by electrons and atomic hydrogen and excited formation are being studied by \citet{Staeuber09}. For the transitions discussed in this section and assuming an excited formation, they find a flux increase by a factor of only $\approx 3$ compared to the peak LTE flux at an excitation temperature of $\approx 30$ K.\\

To constrain rotational temperatures, transitions for different upper levels are necessary. However, only tentative detections of \coplus in YSOs for $J_{\rm up} \geq 14$ are reported (\citealt{Ceccarelli98}). We will thus follow \citet{Black98} and assume $T_{\rm ex}$=30 K, slightly higher than \citet{Staeuber07} (20 K) or \citet{Fuente06} (10 K). For an upper energy level $E_{u}$, the level population is proportional to $e^{-E_{u}/k T_{\rm ex}}/Q(T_{\rm ex})$, with $Q(T)$ being the partition function. Using $T_{\rm ex}=30$~ K, the population of the $J=3$ levels is approximately maximized. The emission of optically thin lines scales with the population of the upper level. Compared to $T_{\rm ex}=30$~ K, the line flux is thus about about $20\%$ weaker for $T_{\rm ex}=20$~ K and a factor of 3.5 for $T_{\rm ex}=10$~ K or $T_{\rm ex}=300$~ K. An excitation temperature of a few hundred K is expected for collisional excitation by electrons and atomic hydrogen (\citealt{Andersson08}). We conclude that a different excitation temperature, within the expected range or assuming an excited formation of CO$^+$, changes the modeled line flux less than an order of magnitude.\\

Synthetic line fluxes are obtained from the solution of the radiative transfer equation. The two lines considered in this work have a very low optical depth with $\tau < 0.08$ for $T_{\rm ex}$=30 K at an assumed line width and molecular column density of 4 km s$^{-1}$ and \tto{9}{12} cm$^{-2}$, respectively. This column density corresponds to an emitting region with a diameter of 60\,000 AU at a density of 10$^{5}$ cm$^{-3}$ with a \coplus fractional abundance of $10^{-10}$. For our application, we can thus neglect self absorption and the integrated flux is obtained from the velocity integrated radiative transfer equation 
\begin{equation} \label{eq:cop_inttemp}
\int T_{\rm mb} \, dv = \frac{h c^3}{8 \pi k} \frac{ g_{u} A_{ul}}{\nu^2} \frac{e^{-E_{u}/k T_{\rm ex}}}{Q(T_{\rm ex})} \int_{\rm LOS} n(s) \, ds \ \ ,
\end{equation}
with $\nu$ being the line frequency, $g_{u}$ the statistical weight of the upper level, $A_{ul}$ the Einstein-A coefficient [s$^{-1}$] and $n(s)$ the \coplus density [cm$^{-3}$]. The integral is taken along the line of sight (LOS). To have comparable results to older literature, we assume a distance of 1 kpc to AFGL 2591, following \citet{Benz07} and \citet{vdTak99}. A larger distance of 1.7 kpc as suggested by \citet{Schneider06} would however not change the main conclusions of this work. The line flux of the spherical 1D model calculated using this method agrees within a few percent with results from \verb!SKY! of the RATRAN radiative transfer code (\citealt{Hogherheijde00}).\\

\begin{figure*}[tbh]
\includegraphics[width=\textwidth]{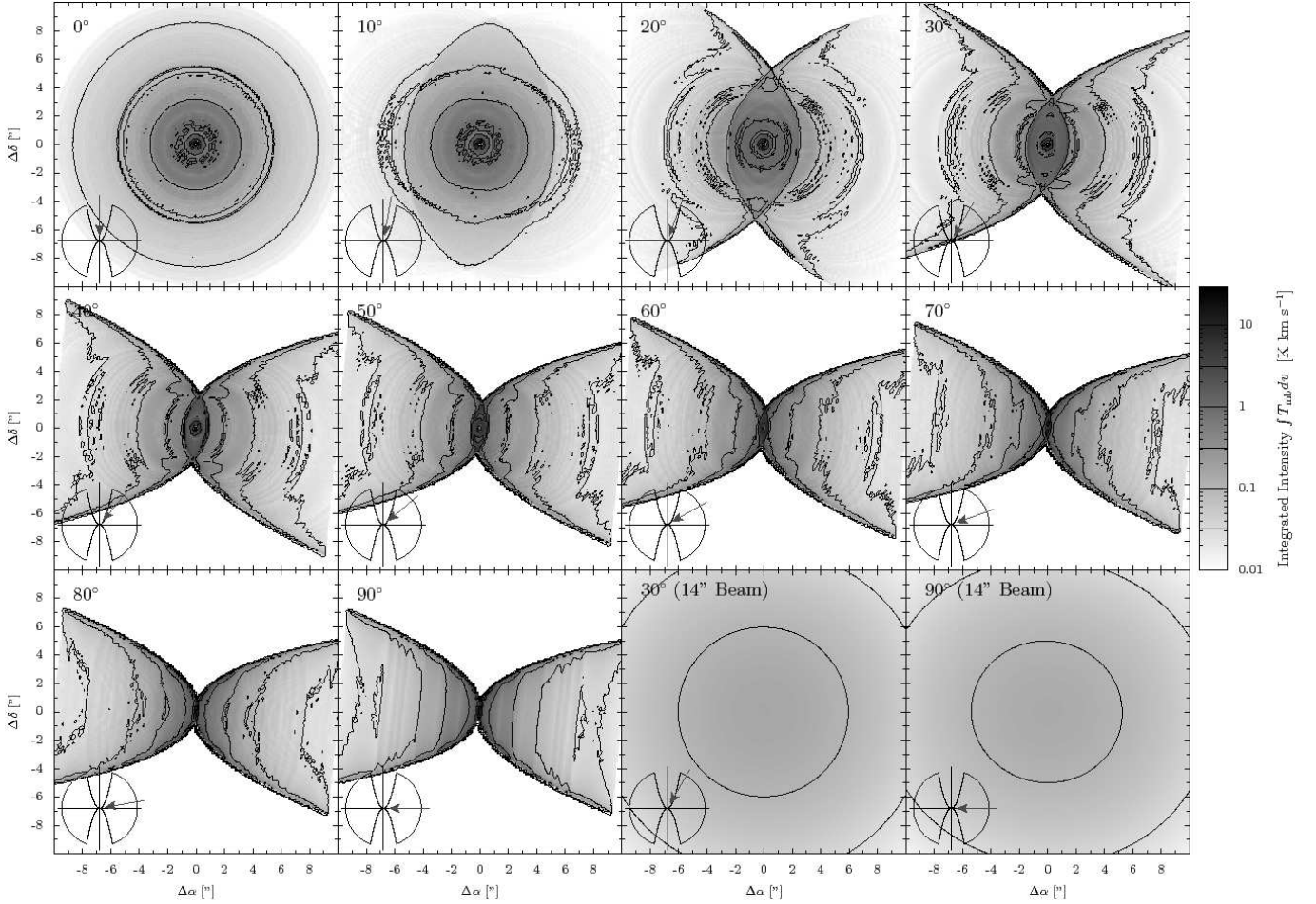}
\caption{Synthetic maps of the velocity integrated \coplus line at 354.014 GHz ($3\frac{7}{2}-2 \frac{5}{2}$) for different inclination angles of the source. The inclination angle is shown with the gray/red arrow onto a symbolic density map in the bottom-left corner of each map. Two plots in the lower right corner give maps convolved the a $14''$ JCMT beam.\label{fig:plot_ray_fine}}
\end{figure*}

Maps of the integrated line flux from the 2D model at different inclination angles are shown in Fig. \ref{fig:plot_ray_fine}. The angular resolution of the images is $0.05''$. We give only results for the $3\frac{7}{2}$ - $2\frac{5}{2}$ transition at 354.014 GHz in this section. The modeled flux of the other fine structure line can be obtained by scaling with 0.7, the ratio of the Einstein-A coefficients and the statistical weights. The modeled maps are convolved with a $14''$ FWHM Gaussian to account for the angular resolution of the JCMT. The effect of different inclination angles is clearly visible in unconvolved maps, but the convolved maps are almost independent of inclination angle.\\

The predicted line fluxes for the JCMT at the center position are given in Table \ref{tab:coplus_sma}. They only vary between 0.16 and 0.14 K km s$^{-1}$ for an inclination of 0$^\circ$ and 90$^\circ$, respectively. This reflects a slightly larger fraction of the outflow walls in the telescope beam for low inclination angles. The difference is however small since most emission is from regions close to the center. Indeed, about between $15\%$ and $30\%$ of the modeled JCMT flux stems from within a radius of $1''$ from the center, corresponding to 1\,000 AU at the adopted distance. An inclination angle of $30^\circ$ was suggested by \citet{vdTak99} and is thus displayed in this and the next section.\\

%
% Figure plot_ray_fine
%

The line flux of the \coplus line at 354.014 GHz as obtained from the spherically symmetric 1D model is with 1.1 $\times$ 10$^{-4}$ K km s$^{-1}$ about 3 orders of magnitude weaker than observed. On the other hand, the results of the axi-symmetric 2D model agree to within a factor of two which we consider as a good agreement given all uncertainties entering the modelling. The $3 \frac{5}{2}$ - $2 \frac{3}{2}$ line at 353.741 GHz however disagrees by about a factor of 5, as the observed line ratio of 1.85 (=[353.741 GHz] / [354.014 GHz]) is larger than the ratio of 0.7 as predicted by the models. Possible explanations are a different excitation mechanism of the fine structure levels or line blending of the 353.741 GHz line, e.g. by the $^{33}{\rm SO}_2$($J_{K_p K_o}=19_{4,16} \rightarrow 19_{3,17}$) line at 353.741 GHz (\citealt{Staeuber07}). To explain the line ratio by a different excitation mechanism, a grossly different excitation temperature would however be required, as discussed earlier in this section.

\subsection{Comparison with SMA observations} \label{sec:compsma}

Interferometric observations of the two \coplus lines in AFGL 2591 have been carried out using the Submillimeter Array (SMA)\footnote{The Submillimeter Array is a joint project between the Smithsonian Astrophysical Observatory and the Academia Sinica Institute of Astronomy and Astrophysics and is funded by the Smithsonian Institution and the Academia Sinica.}. \citet{Benz07} have used the extended configuration of the array with projected baselines covering the range between 38 - 214 k$\lambda$ (32.2 m - 181.2 m). New observations have been carried out in the compact configuration on April 14 2006. These data cover a projected baseline range of 10.6 - 82 k$\lambda$ (9 m - 69.4 m). The frequency setting is the same as for the extended array observation, covering the range between 342.6-344.6 GHz in the LSB and 352.6-354.6 GHz in the USB. Data of this observation will be discussed in a paper in preparation (Bruderer et al. 2009).\\

The RMS noise per velocity bin of 0.5 km s$^{-1}$ is 0.46 Jy Beam$^{-1}$ (extended array), 0.21 Jy Beam$^{-1}$ (compact array) and 0.18 Jy Beam$^{-1}$ when data of both arrays are combined using natural weighting. The better weather conditions and shorter baselines result in a higher sensitivity of the compact array observations. The detection limit for the integrated line can be estimated from $1 \sigma = 1.2 \sqrt{\delta {\rm V} \cdot \delta {\rm v}}\, T_{\rm rms}$, where $T_{\rm rms}$ is the RMS noise per frequency bin, $\delta {\rm V} \approx 4$~ km s$^{-1}$ the expected linewidth and $\delta {\rm v}=0.5$~ km s$^{-1}$ the velocity resolution (\citealt{Maret04}). The factor 1.2 accounts for the uncertainty of the absolute calibration of about $20\%$. For a 3$\sigma$ detection, an integrated flux of 2.3 Jy  km s$^{-1}$ (extended array) and 0.91 Jy Beam$^{-1}$ km s$^{-1}$  (compact and extended arrays combined) is thus necessary.\\

The \coplus line at 354.014 GHz has not been detected at a 3$\sigma$ level in either configurations of the SMA. The 353.741 GHz line is not detected in the extended configuration observations. In the data of the combined array however, a line is found at approximately 353.741 GHz. Is the \coplus line blended by $^{33}{\rm SO}_2$($J_{K_p K_o}=19_{4,16} \rightarrow 19_{3,17}$) at the same frequency? Several other lines of $^{33}{\rm SO}_2$ are within the observed frequency setting, but are not detected despite their presumably higher intensity due to a larger Einstein-A coefficient, lower critical density and upper level energy. \coplus is thus more likely blended by an unidentified line. The blended line has a velocity integrated line flux of 4.9 Jy Beam$^{-1}$ km s$^{-1}$ corresponding to a $25\sigma$ detection. For the predicted line ratio  ([353.741 GHz] / [354.014 GHz]) of 0.7 and the ratio of 1.85 as observed by the JCMT, the \coplus line at 354.014 GHz should have been detected. We thus conclude \coplus is not detected by the SMA observations.\\

\begin{figure}[ht]
\plotone{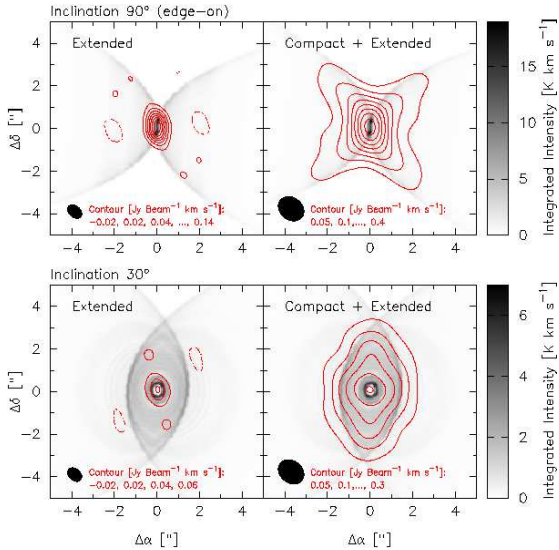}
\caption{Simulated SMA observations [Jy Beam$^{-1}$ km s$^{-1}$] (gray/red contour lines) overlaid on modeled velocity integrated maps [K km s$^{-1}$] (black/white map). Negative fluxes are shown by dashed contour lines. The half-power beam is given in the bottom left corner. For comparison, a 3$\sigma$ detection with the SMA requires a flux of 2.3 (only extended array) / 0.91 (compact+extended array) Jy Beam$^{-1}$ km s$^{-1}$.\label{fig:sma_obsmodel}}
\end{figure}

Is this non-detection consistent with the modeled line flux of the 2D model? To answer that question, synthetic maps (Fig. \ref{fig:plot_ray_fine}) are converted to visibility amplitudes for the $(u,v)$ coverage of the SMA observation using the MIRIAD\footnote{\anchor{http://sma-www.cfa.harvard.edu/miriadWWW/manuals/manuals.html}{http://sma-www.cfa.harvard.edu/miriadWWW/manuals/ manuals.html}} task \verb!uvmodel!. The simulated visibilities are then reduced in the same way as SMA observations: they are inverted and the clean algorithm is applied on the resulting maps. Simulated velocity integrated maps for inclination angles of 30$^\circ$ and 90$^\circ$ are presented in Fig. \ref{fig:sma_obsmodel}. The maximum flux of the simulated maps for the extended array is 0.17 Jy Beam$^{-1}$ km s$^{-1}$ at an inclination of $70^\circ$. For the (u,v) coverage of the combined array, the peak flux is 0.47 Jy Beam$^{-1}$ km s$^{-1}$ at an inclination angle of $60^\circ$. These integrated fluxes correspond to less than $1\sigma$ and the non-detection is thus consistent with the models. As the simulated SMA maps in Fig. \ref{fig:sma_obsmodel}
 show, this non-detection is explained by the limited sensitivity of our observations rather than by over-resolving effects of the interferometer.\\

\begin{table*}
\begin{center}
\caption{Modeled and observed maximum velocity integrated fluxes of the \coplus $3 \frac{7}{2} - 2 \frac{5}{2}$ line at 354.014 GHz.\label{tab:coplus_sma}}
\centering
\begin{tabular}{cccc}
\tableline\tableline
Inclination  & SMA (Ext.\tablenotemark{a})   & SMA (Both.\tablenotemark{b}) & JCMT\\
$[{}^{\circ}]$   & $[$Jy Beam$^{-1}$ km s$^{-1}]$ & $[$Jy Beam$^{-1}$ km s$^{-1}]$ & $[$K km s$^{-1}]$ \\
\tableline
\multicolumn{4}{l}{Axi-symmetric 2D model:} \\
0   & 0.05 & 0.23 & 0.16\\
10  & 0.06 & 0.25 & 0.16\\
20  & 0.05 & 0.26 & 0.16\\
30  & 0.06 & 0.30 & 0.15\\
40  & 0.06 & 0.38 & 0.15\\
50  & 0.11 & 0.44 & 0.15\\
60  & 0.16 & 0.47 & 0.14\\
70  & 0.17 & 0.45 & 0.14\\
80  & 0.16 & 0.44 & 0.14\\
90  & 0.16 & 0.43 & 0.14\\
\tableline
\multicolumn{4}{l}{Spherical 1D model:} \\
 & $< 0.01$ & $< 0.01$ & $1.1 \cdot 10^{-4}$ \\
\tableline
Observed: & $<2.3$\tablenotemark{c} & $<0.91$\tablenotemark{c} & 0.27\tablenotemark{d}\\
\tableline 
\end{tabular}
\tablenotetext{a}{extended configuration}
\tablenotetext{b}{compact configuration}
\tablenotetext{c}{Not detected: $3\sigma$ upper limit}
\tablenotetext{c}{Detected}
\end{center}
\end{table*}

%
% Figure plot_inclination
%

%
% Sec: Conclusions
%
\section{Conclusions and outlook} \label{sec:conclusion}

We have used the grid of chemical models introduced by \citet{Bruderer09a} to construct a detailed two-dimensional model of the high-mass star forming region AFGL 2591. The spherically symmetric 1D models by \citet{Doty02} and \citet{Staeuber04,Staeuber05} have been extended with a low-density outflow region allowing protostellar FUV radiation to escape from the innermost region. In the 2D model, FUV irradiates a larger surface and thus a larger volume compared to spherically symmetric models. The model is used to simulate the line fluxes of CO$^+$ as a prototypical FUV tracer. The main conclusions of this work are:

\begin{enumerate}
\item The existence of concave outflow cavity walls allows the efficient, long-range streaming of FUV radiation to large distances from the central source (Sect. \ref{sec:outflowwalls}).
\item A thin FUV-enhanced layer in the outflow is produced, having a thickness of a few hundreds of AU, and $G_0 >$ 100  ISRF (Sect. \ref{sec:outflowwalls}).
\item A large mass of FUV irradiated material -- more than a solar mass -- can reside in the outflow walls. This combined with the large extent can lead to a situation where the molecular emission from the outflow walls dominates the total line flux in single-dish observations (Sect. \ref{sec:outflowwalls}, Sect. \ref{sec:compobs})
\item Detailed 2D modelling of \coplus including these outflow walls is consistent with single-dish (Sect. \ref{sec:compobs}) and interferometric observations (Sect. \ref{sec:compsma}), while the flux in a 1D model is orders of magnitude too low. This indicates the need for multidimensional chemical models for the interpretation of FUV sensitive molecules.
\item The strong gradients in density and FUV flux require an accurate calculation of the gas temperature structure and inclusion of scattering of FUV radiation in the outflow walls, which can extend the width of the FUV enhanced region significantly (Sect. \ref{sec:outflowwalls}).
\item A region of high FUV ($G_0 >$ 1 ISRF) and low temperature ($T <$ 100 K) leads to the strong possibility of ice mantle processing by FUV photons in the outflow walls (Sect. \ref{sec:outflowwalls}).
\item The abundance of \coplus is also enhanced in a mixing layer between the ionized and warm outflow and the envelope. Mixing alone can however not explain the observed amount of \coplus (Sect. \ref{sec:coplusabundance}, Appendix \ref{sec:coplusmix}).
\end{enumerate}

This paper shows the application of the grid of chemical models (\citealt{Bruderer09a}) for the construction of a multidimensional chemical model of a YSO  envelope. This interpolation method simplifies the construction of two or three dimensional chemical models considerably and the gain in speed allows the self-consistent calculation of the temperature structure with the chemical abundances (Sect. \ref{sec:temp2dafgl}). Our work shows that such detailed modelling including the influence of high-energy irradiation and the geometrical shape of the envelope can be necessary even to explain spatially unresolved single-dish observations.\\

Some hydrides (e.g. SH$^+$ or NH$^+$) will be observable for the first time using the upcoming HIFI spectrometer onboard Herschel. They are expected to be chemical tracers of FUV/X-ray radiation (\citealt{Staeuber04,Staeuber05,Staeuber07}) and further study will be carried out to investigate the influence of the geometry on these species for the interpretation of the Herschel data. Direct imaging of the outflow region will be available with the Atacama Large Millimeter/submillimeter Array (ALMA). These high spatial resolution observations will deliver additional constraints on the geometry.

\acknowledgments

We thank Kaspar Arzner, Pascal St\"auber and Jes J\o rgensen for useful discussions and an anonymus referee for his/her valuable comments. Michiel Hogerheijde and Floris van der Tak are acknowledged for use of their RATRAN code. This work was partially supported under grants from The Research Corporation and the NASA grant NNX08AH28G (SDD). Astrochemistry in Leiden is supported by the Netherlands Research School for Astronomy (NOVA) and by a Spinoza grant from the Netherlands Organization for Scientific Research (NWO). The submillimeter work at ETH is supported by the Swiss National Science Foundation grant 200020-113556.\\

\noindent
{\it Facilities:} \facility{SMA}, \facility{JCMT}

\clearpage

\appendix

\section{A. A Monte Carlo code for FUV radiative transfer} \label{sec:mc_uvimplement}

In the first appendix, we discuss the implementation of the radiative transfer calculation of the FUV radiation. Unlike the work of \citet{vanZadelhoff03}, our calculation is implemented on a three dimensional Cartesian grid and thus already able the solve three dimensional problems. To improve convergence, points with constant $z$ and $r=\sqrt{x^2+y^2}$ are averaged for the axi-symmetric model in this work. To save memory, only cells with positve values of $x$, $y$ and $z$ are stored, but photon packages can of course still travel from one quadrant to another. The flow of the calculation is 
\begin{enumerate}
\item A photon package is started from the FUV source at an arbitrary direction with a luminosity of $L_{\rm package}=L_{\rm source}/(8 \cdot N_{\rm photon})$ [erg s$^{-1}$], where $L_{\rm source}$ is the luminosity of the FUV source in the 6 - 13.6 eV band. $N_{\rm phton}$ is the number of photon packages in the simulation run. The factor 8 takes into account that only cells with positive values of $x$, $y$ and $z$ are stored.
\item As the photon package travels along a straight line, $L_{\rm package}$ is attenuated by $\exp\left(-\tau_{\rm loc}\right)$ with $\tau_{\rm loc}=\sigma_{\rm ext}\,n(x,y,z)\,\Delta s$, the local FUV optical depth. The optical depth depends on $\sigma_{\rm ext}$, the extinction cross-section, $n(x,y,z)$ the local density and $\Delta s$ the step size of the code. The total extinction, that a photon package suffers from the source to the current position $\tau_{\rm tot} = \sum \tau_{\rm loc}$ is stored. To avoid discretization errors, especially close to the source, we use the analytical expression for the local density.
\item If the photon package passes the border of a cell, the local intensity is updated. The attenuated intensity is obtained from 
\[
I_{\rm att}(x,y,z) = \sum L_{\rm package} \times \frac{\left| \vec{n}_{\rm photon} \cdot \vec{n}_{\rm cell}\right|}{A} \ \ ,
\]
where the sum is taken over all photon packages passing the cell. The surface of the cell wall which the photon package just passed is given by $A$ [cm$^2$] and the directions of the photon package and the normal of the cell separation is given by $\vec{n}_{\rm photon}$ and $\vec{n}_{\rm cell}$, respectively.
\item The unattenuated intensity $I_{\rm unatt}$~ is calculated using the same expression as $I_{\rm att}$~, except $L_{\rm package}$~ is replaced by $L_{\rm package}\times \exp\left(+\tau_{\rm tot}\right)$.
\item If a random number [0,1] is larger than $\exp\left(-\tau_{\rm scat}\right)$, with $\tau_{\rm scat} = \sigma_{\rm ext}\,n(x,y,z)\,\Delta s$, the scattering-optical depth, a new direction of the photon package is chosen following the implemented phase function $\phi(\Delta)$. For our work, we use the tabulated function of \citet{Draine03a} for an average Milky Way dust with $R_V = 3.1$~ and C/H=56 ppm in PAHs.
\item For each photon package, steps 2-5 are repeated, until $L_{\rm package}$ is smaller then a certain threshold or the package left the simulated area.
\item The code propagates 10$^5$ photon packages and then averages for the attenuated intensity $I_{\rm att}$ in the axi-symmetric model. Three subsequent solutions are used to calculate the signal-to-noise ratio (SNR), defined by
\[
{\rm SNR}^{-1} \equiv \max_{i=1,3}\left( \frac{I^i_{\rm att}-\langle I^i_{\rm att} \rangle}{\langle I^i_{\rm att} \rangle} \right) \ \ .
\]
Only cells with an averaged intensity larger than 10$^{-4}$~ ISRF are considered.  Convergence is reached, if the SNR exceeds 50 in all cells. This means less than $2\%$ difference to the average. The application in this work requires \tto{9}{6} photon packages to achieve this accuracy.
\end{enumerate}

As a benchmark test, the code is re-run for the density distribution in Sect. \ref{sec:dens2dafgl}, but with scattering switched off. In this way, an analytical solution can be compared to the result of the code. For the analytical solution, only the distance $r$ and column density $N({\rm H})$ [cm$^{-2}$] to the FUV source are needed to derive the local FUV flux [erg s$^{-1}$ cm$^{-2}$]
\[
F^{\rm loc}=\frac{L_{\rm UV}}{4 \pi r^2} \times \exp\left(-\tau_{\rm UV}\right) \,
\]
with the FUV luminosity $L_{\rm UV} =4.2 \times 10^{37}$ erg s$^{-1}$ and the FUV optical depth $\tau_{\rm UV} = 2.4 (N({\rm H})\, {\rm [cm}^{-2}{\rm ]}) / 1.87 \times 10^{21}$. The conversion factor of 1 ISRF $\equiv$ \tto{1.6}{-3} erg s$^{-1}$ cm$^{-2}$ yields the flux $G_0$ in units [ISRF]. The agreement between the code and the analytical calculation is tested for points with $G_0 > 10^{-2}$. A very good agreement is found, with deviations less than $50\%$.\\

\section{B. Thermal balance calculation}  \label{sec:app_thermal}

In this appendix, references for the heating and cooling rates used in Sect. \ref{sec:temp2dafgl} are given and the temperature balance calculation is benchmarked with PDR models.\\

\subsection*{Heating rates}

\textit{Photoelectric heating:} The heating rate of photoelectrons from small dust grains and PAHs is $\Gamma_{PE} = 10^{-24} \, \epsilon \,n\, G_0$~ erg cm$^{-3}$ s$^{-1}$, with the density $n$, the attenuated FUV field $G_0$ and an efficiency $\epsilon$. The efficiency depends on the ionization fraction of the grains, since the work function increases for ionized dust grains. \citet{Bakes94} give an analytical fit for $\epsilon$ as a function of the ratio $\gamma$ between ionization and recombination rate ($\gamma=G_0 \sqrt{T} / n_e$, with the electron density $n_e$ and the temperature $T$).\\

\noindent
\textit{\hh heating:} Different heating processes involving molecular hydrogen are considered: (i) Collisional de-excitation of vibrationally pumped \hh by FUV photons (FUV-pumping). (ii) Formation heating: \hh forming on dust releases part of the binding energy to the gas. (iii) Photodissociation of \hh heats the gas. The rates for process (i) and (iii) depend on the local FUV field. Since line absorption is responsible for the pumping and dissociation of the molecule, we reduce the local FUV field by the self shielding factor $\beta(\tau)$ (\citealt{Draine96}). Rates for (i) and (ii) are implemented by the analytical expression of \citet{Rollig06}. The rate of process (iii) is taken from \citet{Meijerink05a}.\\

\noindent
\textit{C ionization:} Photoionization of atomic carbon, C + $\gamma_{\rm UV}$ $\rightarrow$  C$^+$  + e$^{-}$, releases an energy of 1.06 eV to the gas and thus contributes to the heating. We follow \citet{Tielens05} for the implementation.\\

\noindent
\textit{X-ray heating:} Fast photoelectrons produced by X-ray photons lose part of their energy through Coulomb interaction with thermal electrons. The heating rate is given by $\Gamma_{\rm X} = \eta \, n \, H_{\rm X}$, where H$_{\rm X}$ [erg s$^{-1}$] is the energy deposition of X-rays in the gas (c.f. \citealt{Staeuber05}) per density. The efficiency factor $\eta$ depends on the H/\hh ratio and the ionization fraction. An analytical fit for $\eta$ to detailed calculations is given in \citet{Dalgarno99}.\\

\noindent
\textit{Cosmic ray heating:} For low degrees of ionization $x < 10^{-4}$, about 9  eV per primary ionization through a cosmic ray particle is used to heat the gas. The heating rate is thus $\Gamma_{\rm cr}= 1.5 \times 10^{-11} \, \zeta_{{\rm H}_2} \, n$~ erg cm$^{-3}$ s$^{-1}$.

\subsection*{Cooling rates}

\noindent
\textit{Atomic fine structure lines:} Forbidden fine structure lines of O, C and C$^+$ are important coolants at the surface of the FUV irradiated zone. We consider the [OI] 63 $\mu$m, [OI] 146 $\mu$m, [CI] 369 $\mu$m,  [CI] 609 $\mu$m and [CII] 158 $\mu$m lines. The cooling rate is obtained from 
\[
\Lambda_{\rm line}=A_{ul} \, h\nu_{ul} \, \beta(\tau_{ul}) \, n_u(X) \ \ ,
\]
with the Einstein-A coefficient $A_{ul}$ [s$^{-1}$] of the transition $u$ $\rightarrow$ $l$, $\nu_{ul}$ [s$^{-1}$] the line frequency, $\tau_{ul}$ the optical depth, $\beta(\tau_{ul})$ the escape probability and $n_u(X)$ [cm$^{-3}$], the density of the species X in the excited level $u$. An escape probability formalism (e.g. \citealt{Tielens05}) is used to calculate the level population iteratively. Collisional excitation by H, \hh and electrons is taken into account - the abundance of the collision partners is read out of the chemical grid. We use the same collision rate coefficients as \citet{Meijerink06a}. The optical depth $\tau$ is calculated using the molecular column density to the outflow cone (Sect. \ref{sec:temp2dafgl}).\\

\noindent
\textit{\hh line cooling:} Vibrational lines of \hh can contribute to the cooling of the gas. Due to the large gap between the ground state and the first excited state, corresponding to about 6\,000 K, we use the two level approximation given in \citet{Rollig06} as a simplification. Again, self shielding is taken into account for the radiative pumping by FUV photons.\\

\noindent
\textit{Gas-grain cooling/heating:} The difference in temperature between $T_{\rm gas}$ and the $T_{\rm dust}$ leads to a transfer of heat. This can be cooling close to the FUV irradiated zone, where $T_{\rm dust} < T_{\rm gas}$, or heating deeper in the envelope. The rates are proportional to $T_{\rm dust} - T_{\rm  gas}$. We implement the results of \citet{Hollenbach89} for a minimal grain size of 10 $\mbox{\AA}$.\\

\noindent
\textit{Cooling by CO and H$_2$O:} Molecules can contribute to the gas cooling by rotational lines at low temperature and ro-vibrational lines at higher temperature. We include line cooling by CO and \hho using the fitted rates of \citet{Neufeld93} (for $T>100$~ K) and \citet{Neufeld95} (for $T \leqslant 100$~ K). The rates depend on the molecular column density, the temperature and the density of the collision partner. Their work considered excitation by H$_2$. Excitation by electrons and atomic hydrogen is taken into account by the approximation given in \citet{Yan97} and \citet{Meijerink05a}. The molecular column density is obtained in the same way as for the cooling through atomic fine structure lines. Cooling by isotopes (${}^{13}$CO, C${}^{18}$O, H$_2^{18}$O) can be important due to the smaller optical depth. Isotope ratios by \citet{Wilson94} are used to scale the molecular density and the column density.\\

\noindent
\textit{Recombination:} Recombination of electrons with grains and PAHs can cool the gas. We implement the fit-results by \citet{Bakes94} for our calculation. This rate also depends on the ratio $\gamma$ between ionization and recombination rate, due to increased Coulomb interaction in gas with a high ionization fraction.\\

\noindent
\textit{Ly$\alpha$ emission:} Atomic hydrogen excited by electron collisions emit Ly$\alpha$ photons and thus contribute to the cooling. This process is only efficient at temperatures higher than a few thousand K. We use the cooling rate given in \citet{Sternberg89}.

\subsection*{Benchmark test}

We compare our calculation of $T_{\rm gas}$ to PDR codes, in a similar situation as model V4 in the PDR comparison study by \citet{Roellig07}. The test consists of a plane parallel slab at a density of 10$^{5.5}$~ cm$^{-3}$ and a FUV irradiation of $\chi = G_0/1.71 = 10^5$. We assume an X-ray flux of about 0.04 erg s$^{-1}$ cm$^{-2}$ consistent to the AFGL 2591 model at $z=1\,000$ AU assuming an X-ray luminosity of $L_{\rm X}$ = 10$^{32}$~ erg s$^{-1}$. Differences in the resulting gas temperature, when X-rays are switched off, are however negligible.\\

\begin{figure*}[tbh]
\includegraphics[width=\textwidth]{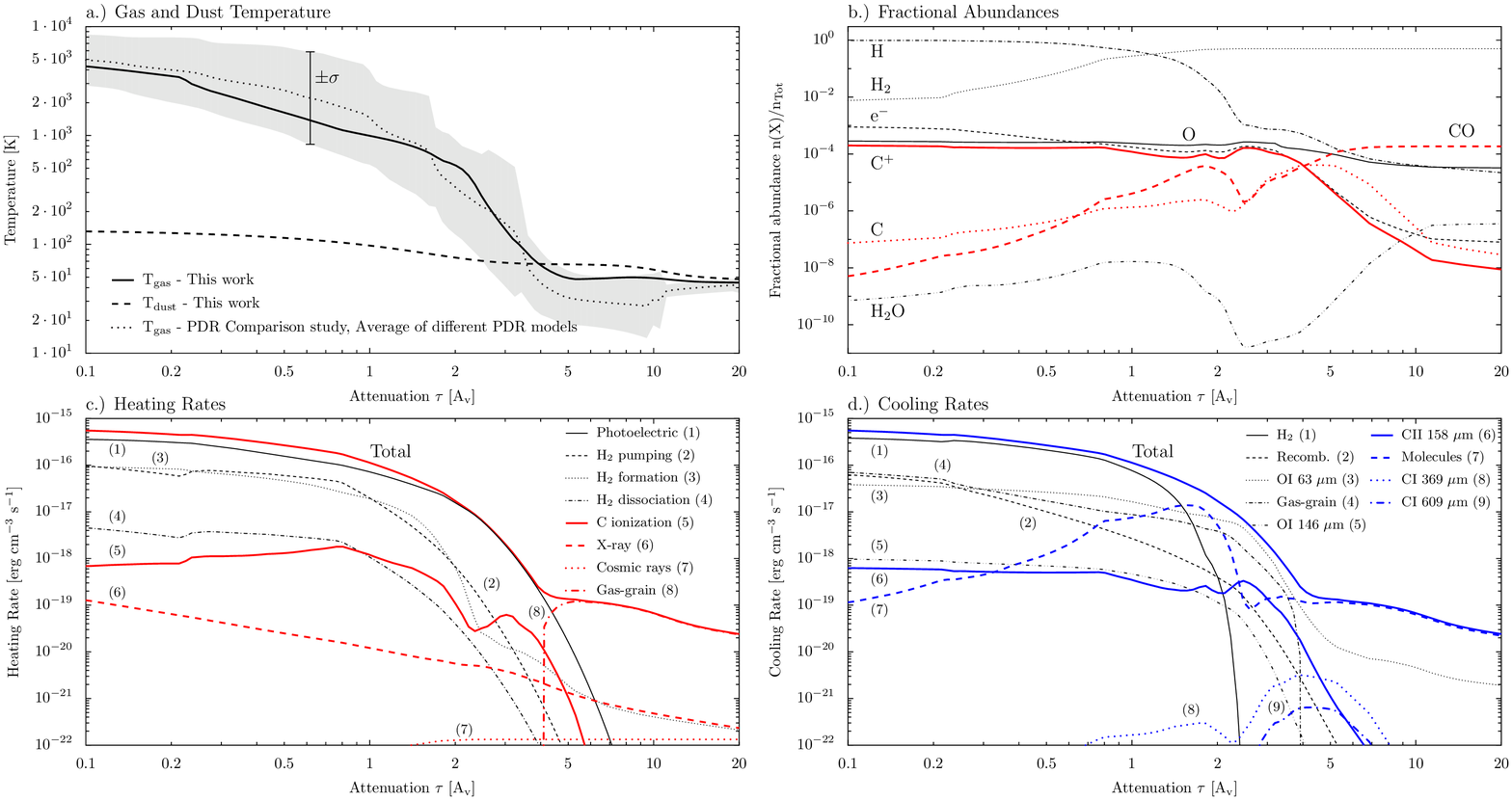}
\caption{Benchmark test of the heating/cooling calculation. \textbf{a.)} Result of the thermal balance calculation described in Sect. \ref{sec:app_thermal}. The gray shaded region gives the range of the PDR comparision study by \citet{Roellig07} (one standard deviation, mean=average=dotted line). \textbf{b.)} Fractional abundances of species relevant to the thermal balance calculation. \textbf{c.)} Heating rates, the line on top gives the total rate. \textbf{d.)} Cooling rates, the line on top gives the total rate.\label{fig:tempbench}}
\end{figure*}

Figure \ref{fig:tempbench}a shows the dust and gas temperature obtained with the thermal balance calculation. As expected, the gas temperature is much higher than the dust temperature for low optical depth, while both temperatures agree well at high optical depth and density (\citealt{Doty97a}). The mean of the gas temperature obtained with different PDR codes is given by a dotted line, along with one standard deviation given by the gray shaded region. The results of our code agrees very well with the results of the PDR codes, while the calculation only takes a few seconds.\\

Figure \ref{fig:tempbench}b gives the abundances of molecules involved in heating or cooling processes. As \citet{Meijerink05a}, we use a ``one-line'' approximation for the \hh and CO self-shielding and are thus not able to reproduce the exact position of the H/\hh transition. Notice the very low water abundance due to the destruction of water by X-rays  (\citealt{Staeuber06}). In Fig. \ref{fig:tempbench}c, the rates of each heating process is given along with the total heating rate. At low optical depth, photoelectric heating and \hh-pumping are the main heating sources, while dust-gas coupling heats the gas at high optical depth ($\tau > 6$ in the example). X-ray heating does not contribute significantly to the heating rate, but the destruction the important coolant \hho by X-rays slightly enhances the temperature. Fig. \ref{fig:tempbench}d shows rates of different cooling processes. At the edge of the calculated region, \hh ro-vibrational lines govern the cooling. [OI] 63 $\mu$m and molecular cooling is important at higher optical depth. Due to the low abundance of water, the main molecular coolant is CO.\\

\section{C. \coplus production in a mixing layer} \label{sec:coplusmix}

\begin{figure}[ht]
\plotone{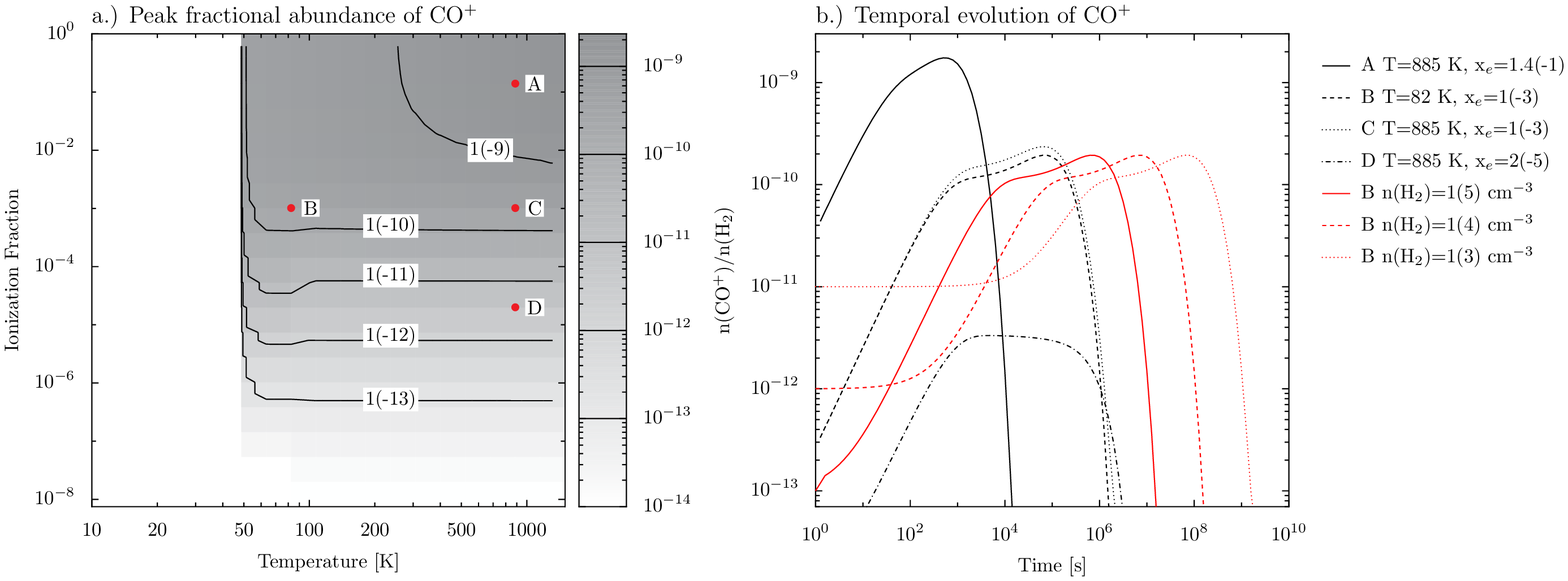}
\caption{\textbf{a.)} Peak fractional abundance of the temporal evolution of CO$^+$ for different gas temperatures and initial ionization fractions is displayed by marked isocontours and gray scale. The density was chosen to be $10^6$ cm$^{-3}$. \textbf{b.)} Temporal evolution of the fractional abundance at points A, B, C, D (black lines). For a temperature of 82 K and an ionization of $10^{-3}$ (point B), densities of $10^3$ cm$^{-3}$, $10^4$ cm$^{-3}$ and $10^5$ cm$^{-3}$ are given in red lines.\label{fig:mixtoy}}
\end{figure}

As a toy-model for the mixing layer between the warm and ionized outflow and the envelope, we assume the outflow material to be in the form of electrons (e$^-$) and ionized hydrogen (H$^+$). The chemical evolution of a parcel of gas consisting of a mixture of outflow and inflow material is modeled similarly to Sect. \ref{sec:chemcoplus}. Free parameters are the temperature and the initial abundances of e$^-$ and H$^+$. The peak fractional abundance of the temporal evolution of \coplus is given in Fig. \ref{fig:mixtoy} together with the temporal evolution of the fractional abundance for selected temperatures and initial ionizations.\\

For temperatures above 60 K, formaldehyde (H$_2$CO) is evaporated from ice mantles. The reaction H$_2$CO + H$^+$ $\rightarrow$ CO$^+$ + H$_2$ + H forms \coplus and it is destroyed by the reaction with H$_2$. For very high initial ionization fractions ($> 10^{-2}$), the dissociative recombination \coplus + e$^{-}$ $\rightarrow$ C + O with rate coefficient $k \propto 1/\sqrt{T}$ becomes important and the peak fractional abundance is weakly temperature dependent. Due to the short recombination time-scale of e$^-$ with H$^+$ and grains, the \coplus abundance peaks at young chemical ages and then decreases quickly.\\

The column density along the mixing layer is $N({\rm CO}^+) \approx L \cdot n \cdot x_{{\rm CO}^+}$, with the width of the mixing layer $L$, the gas density $n$ and the peak fractional abundance $x_{{\rm CO}^+}$. We set $L = v_A \cdot t$, with the Alfv\'en velocity $v_A \leq 10$ km s$^{-1}$ and the time $t$, defined by the point of evolution where the abundance of \coplus has dropped by 1 order of magnitude compared to the peak. The peak fractional abundance $x_{{\rm CO}^+}$ and the product $t \cdot n$ are approximately independent of density as Fig. \ref{fig:mixtoy}b shows. On the other hand, $t \cdot x_{{\rm CO}^+}$ is roughly independent of temperature and peaks at an electron fraction of about $10^{-3}$, where $t \approx 5 \times 10^5$ s and $x_{{\rm CO}^+} \approx 2 \times 10^{-10}$ for a density of $10^6$ cm$^{-3}$. We obtain for the upper limit of the column denisty due to mixing $N({\rm CO}^+) \approx v_A \cdot t \cdot n \cdot x_{{\rm CO}^+} \approx 10^8$ cm$^{-2}$. In the FUV model of Sect. \ref{sec:afgl2dmod}, the column density of \coplus along constant $z$ is of order $10^{10}$ cm$^{-2}$. We conclude that mixing alone cannot reproduce the observed amount of CO$^+$.

\bibliographystyle{apj,apj-jour}

\clearpage

\end{document}